\documentclass[%
 reprint,
superscriptaddress,
showpacs,
showkeys,
nofootinbib,
 amsmath,amssymb,
 aps,
]{revtex4-1}

\usepackage{graphicx}
\usepackage{dcolumn}
\usepackage{bm}
\usepackage{amsthm} 
\usepackage{subfigure} 
\usepackage{hyperref}

\newcommand{\rme}{\mathrm{e}}

\input{Qcircuit}

\begin{document}


\title{A State Distillation Protocol to Implement Arbitrary Single-qubit Rotations}

\author{Guillaume Duclos-Cianci}
\email{Guillaume.Duclos-Cianci@USherbrooke.ca}
\affiliation{
D\'{e}partment de Physique, Universit\'{e} de Sherbrooke, Sherbrooke, Qu\'{e}bec, J1K 2R1 (Canada)
}
\author{Krysta M.~Svore}
\email{ksvore@microsoft.com}
\affiliation{%
 Quantum Architectures and Computation Group, Microsoft Research, Redmond, WA 98052 (USA)
}%

\date{\today}

\begin{abstract}
An important task required to build a scalable, fault-tolerant quantum computer is to efficiently represent an arbitrary single-qubit rotation by fault-tolerant quantum operations.
Traditionally, the method for decomposing a single-qubit unitary into a discrete set of gates is Solovay-Kitaev decomposition, which in practice produces a sequence of depth $O(\log^c(1/\epsilon))$, where $c\sim3.97$ is the state-of-the-art.
The proven lower bound is $c=1$, however an efficient algorithm that saturates this bound is unknown.
In this paper, we present an alternative to Solovay-Kitaev decomposition employing state distillation techniques which reduces $c$ to between $1.12$ and $2.27$, depending on the setting.
For a given single-qubit rotation, our protocol significantly lowers the length of the approximating sequence and the number of required resource states (ancillary qubits). In addition, our protocol is robust to noise in the resource states.

\end{abstract}

\pacs{03.67.Lx, 03.67.Pp, 03.65.Fd}
\keywords{state distillation, Solovay-Kitaev decomposition}

\maketitle

Given recent progress in quantum algorithms, quantum error correction, and quantum hardware, a \emph{scalable} quantum computer is becoming closer and closer to reality.
For many proposed quantum computer architectures, e.g., topological systems based on the braiding of non-Abelian anyons \cite{Kitaev03,Freedman98,Preskill98,FLW02,Nayak08} or the surface code model based on code deformation \cite{BM09,F09}, so-called Clifford operations and stabilizer state preparations or measurements can be implemented efficiently and accurately.
However, these operations alone are not sufficient for quantum universality since they can be simulated classically \cite{G98,AG04,NC00}.
One technique to achieve quantum universality is to use magic state distillation \cite{BK05,MEK05,BH12} to augment the set with a single non-Clifford operation, e.g., the single-qubit $\pi/8$ gate, $T$.
This augmented set can be used to approximate any single-qubit unitary using Solovay-Kitaev decomposition \cite{DN05}.

The Solovay-Kitaev theorem \cite{KitaevEtAl2002} states that for any $\epsilon$ and single-qubit gate $U$,
 $U$ can be approximated to precision $\epsilon$ using $\Theta(\log^c(1/\epsilon))$ gates drawn from a universal, discrete gate set, where $c$ is a small constant.
State-of-the-art implementations of Solovay-Kitaev decomposition result in $c\sim3.97$ \cite{BS12,DN05}, resulting in an average decomposition sequence with hundreds to thousands of $T$ gates \cite{BS12}.
Each $T$ gate requires a number of copies of a quantum magic state $\ket{H}$ (Hadamard +1-eigenstate), where the number depends on the specific state distillation protocol and purity of the state \cite{BK05,MEK05,BH12}.
Therefore, it is especially important to minimize the number of $T$ gates when decomposing a rotation, since each $T$ gate requires additional ancillary qubits for implementation.

In this paper, we present an alternative protocol to Solovay-Kitaev decomposition that allows the implementation of single-qubit rotations through the use of distilled magic $\ket{H}$ states.
We show that the resources required by our protocol are substantially fewer than the resources required by state-of-the-art implementations of Solovay-Kitaev decomposition \cite{BS12}, in both the number of gates and the number of quantum magic states necessary to apply an arbitrary single-qubit unitary.

\section{Magic State Distillation and Implementation of Rotations}

Solovay-Kitaev decomposition \cite{DN05,KitaevEtAl2002,BS12} enables the approximation of any gate using an approximately universal set of elementary gates, e.g., $\{H,T,\Lambda{(X)}\}$, where $\Lambda{(X)}$ denotes the controlled-{\tt NOT} gate.
In particular, one can approximate any single-qubit unitary operation using the set $\{H,T\}$.
Magic state distillation is then used to produce the magic $\ket H$ states necessary to implement the $T$ gate.

We call a state $\ket{\psi}$ \emph{magic} if given $n$ noisy copies of $\ket{\psi}$ and the ability to perform perfect Clifford operations, we can obtain, or ``distill", a purer copy of $\ket{\psi}$ from a Clifford circuit applied to the $n$ noisy copies of $\ket{\psi}$.
We can then obtain even purer states which are arbitrarily close to the perfect state by applying the protocol recursively \cite{BK05,MEK05,BH12}.
These distilled states can be used to implement non-Clifford operations, e.g., the $T$ gate, as described below.

We briefly review how to perform an arbitrary rotation about the $Z$-axis using a resource state, and how to apply the $T$ gate.
We assume that Clifford operations are applied perfectly, since they can be implemented fault-tolerantly, and that the resource states are very close to pure.
One can use the protocols of \cite{BK05,MEK05} to perform the initial distillation in order to obtain arbitrarily pure resource states for input.
We focus on the $+1$ eigenstate of the Hadamard operation, $H$,
\begin{eqnarray*}
  \ket{H} &=& \cos \frac\pi8\ket0 + \sin \frac\pi8 \ket1,
\end{eqnarray*}
and omit the cost of the initial distillation of the $\ket{H}$ state.
We concentrate on single-qubit states found in either the $XZ$- or $XY$-plane of the Bloch sphere.
Note that one can easily move a state from one plane to the other by the application of the Clifford $HSHX$ operation.


Suppose we start with states $\ket{Z(\theta)}$ and $\ket\psi$:
\begin{eqnarray*}
  \ket{Z(\theta)} &=& \ket0 + \rme^{i\theta}\ket1, \\
  \ket{\psi} &=& a\ket0 + b\ket1.
\end{eqnarray*}

\begin{figure}
\[\Qcircuit @C=1em @R=1em {
\lstick{\ket{Z(\theta)}} & \gate{X (Z)}  & \meter & \rstick{\ket{m}} \cw \\
\lstick{\ket{\psi}}      & \ctrl{-1} & \rstick{Z (X) (-1^m \theta)\ket{\psi}} \qw \\
}\]
\caption{Circuit randomly implementing a rotation of angle $\pm\theta$ around the $Z(X)$-axis.
}
\label{fig:RotCircs}
\end{figure}
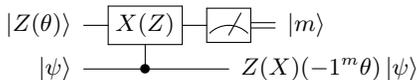

The circuit to implement a rotation around the $Z$-axis, presented in Fig.~\ref{fig:RotCircs}, leads to the two-qubit state
\begin{eqnarray*}
  \ket{Z(\theta)}\ket\psi &=& a\ket{00} + b\ket{01} + a\rme^{i\theta}\ket{10} + b\rme^{i\theta}\ket{11}\\
  &\xrightarrow{\Lambda{(X)}}& a\ket{00} + b\ket{11} + a\rme^{i\theta}\ket{10} + b\rme^{i\theta}\ket{01}.
\end{eqnarray*}

Upon measurement of the first qubit in the computational basis, we obtain
\begin{eqnarray*}
& \xrightarrow{m=0} &  a\ket{0} + b\rme^{i\theta}\ket{1},\\
& \xrightarrow{m=1} &  a\rme^{i\theta}\ket{0} + b\ket{1} = a\ket{0} + b\rme^{-i\theta}\ket{1},
\end{eqnarray*}
each with probability $1/2$.
Thus, the angle of rotation is chosen at random to be $\theta$ or $-\theta$, up to global phase.
An analogous circuit performs a rotation about the $X$-axis.
Similar circuits can be found in \cite{F09}.

%

As an important example of this procedure, consider the $XY$-plane version of the $\ket H$ state:
\begin{equation*}
\ket{Z(\pi/4)} = HSHX \ket H = \ket0+\rme^{i\pi/4}\ket 1.
\end{equation*}
Using the circuit in Figure \ref{fig:RotCircs}, we can implement a $Z$-rotation of angle $\pm\pi/4$, producing at random either the $T$ gate or its adjoint, $T^\dagger$.
In this particular case, we can deterministically apply the desired gate $T$ or $T^\dagger$ by applying the phase gate $S$, since $ST^\dagger = T$.
In general, however, this deterministic correction will not be possible.


\section{New states from $\ket H$ states}

In this section, we show that we can use a very simple two-qubit Clifford circuit to obtain other non-stabilizer states using only $\ket H$ states as an initial resource, and then show that these states enable the approximation of any single-qubit rotation.
We assume that we are provided with perfect copies of $\ket H$.
We would like to minimize the number of $\ket H$ states required to implement an arbitrary single-qubit rotation, since these distilled states can be costly to produce.

\begin{figure}
\[
\Qcircuit @C=1em @R=1em {
\lstick{\ket{H_0}} & \gate{X}  & \meter & \rstick{\ket{0} (\ket{1})} \cw \\
\lstick{\ket{H_i}} & \ctrl{-1} & \rstick{\ket{H_{i+1}} (\ket{H_{i-1}})} \qw \\
}\]
\caption{Two-qubit circuit used to obtain new $\ket{H_i}$ states from initial resource states $\ket{H_0}$. Upon measuring the 0 outcome, the output state is $\ket{H_{i+1}}$. Upon measuring the 1 outcome, the output state is $\ket{H_{i-1}}$.}\label{fig:2QbLadderCircs}
\end{figure}
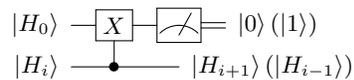
%

Consider the circuit of Fig.~\ref{fig:2QbLadderCircs}. One can easily verify that it measures the parity of the two input qubits and decodes the resulting state into the second qubit.
We begin by considering the two inputs to be $\ket H$ states.
We define $\theta_0=\frac\pi8$ and $\ket H = \ket{H_0} = \cos \theta_0 \ket 0 + \sin \theta_0 \ket 1$.
The circuit begins as:
\begin{eqnarray*}
  \ket{H_0}\ket{H_0} &=& \cos^2 \theta_0 \ket{00} + \sin^2 \theta_0 \ket{11}\\
  & & + \cos \theta_0 \sin \theta_0 (\ket{01} + \ket{10})\\
  & \xrightarrow{\Lambda{(X)}} & \cos^2 \theta_0 \ket{00} + \sin^2 \theta_0 \ket{01}\\
  & & + \cos \theta_0 \sin \theta_0 (\ket{11} + \ket{10}).
\end{eqnarray*}

Upon measurement of the first qubit, we have
\begin{eqnarray*}
  & \xrightarrow{m=0}&  \frac{\cos^2\theta_0\ket0 + \sin^2\theta_0\ket1}{\cos^4\theta_0+\sin^4\theta_0},\\
  & \xrightarrow{m=1}&  \frac{1}{\sqrt2}(\ket0 + \ket1).
\end{eqnarray*}

We define $\theta_1$ such that
\begin{eqnarray*}
  \cos\theta_1\ket0 + \sin\theta_1\ket1& = &  \frac{\cos\theta_0\ket0 + \sin\theta_0\ket1}{\cos^4\theta_0+\sin^4\theta_0},
\end{eqnarray*}
from which we deduce $\cot\theta_1=\cot^2\theta_0$. We define $\ket{H_1}=\cos\theta_1\ket0 + \sin\theta_1\ket1$, a non-stabilizer state obtained from $\ket H$ states, Clifford operations, and measurements. If the outcome of the measurement is 1, then we obtain a stabilizer state
and discard the output (see Fig.~\ref{fig:2QbLadderCircs}).

The two measurement outcomes occur with respective probabilities
\begin{eqnarray*}
  p_0 &=& \cos^4 \theta_0 +\sin^4 \theta_0 =\frac{3}{4},\\
  p_1 &=& 1-p_0 = \frac{1}{4}.
\end{eqnarray*}

We now recurse on this protocol using the non-stabilizer states produced by the previous round of the protocol as part of the input to the circuit of Fig.~\ref{fig:2QbLadderCircs}. We define
\begin{eqnarray*}
  \ket{H_i} &=& \cos \theta_i \ket0 + \sin \theta_i \ket1, \\
  \cot \theta_i &=& \cot^{i+1} \theta_0.
\end{eqnarray*}
If we use as input a copy of $\ket{H_i}$ and a copy of $\ket{H_0}$, we have
\begin{eqnarray*}
  \ket{H_0}\ket{H_i} &=& \cos \theta_0 \cos \theta_i \ket{00} + \sin \theta_0 \sin \theta_i \ket{11}\\
  & & + \sin \theta_0 \cos \theta_i \ket{10} +\cos \theta_0  \sin \theta_i  \ket{01},\\
  & \xrightarrow{\Lambda{(X)}} & \cos \theta_0 \cos \theta_i \ket{00} + \sin \theta_0 \sin \theta_i \ket{01}\\
  & & + \sin \theta_0 \cos \theta_i \ket{10} + \cos \theta_0 \sin \theta_i \ket{11}.
\end{eqnarray*}
Upon measurement of the first qubit, we have
\begin{eqnarray*}
  &\xrightarrow{m=0}&  (\cos \theta' \ket0 + \sin \theta' \ket1),\\
  &\xrightarrow{m=1}&  (\cos \theta'' \ket0 + \sin \theta'' \ket1),
\end{eqnarray*}
where
\begin{eqnarray*}
  \cot \theta'  &=& \cot \theta_i \cot \theta_0 =\cot^{i+2} \theta_0 =\cot \theta_{i+1},\\
  \cot \theta''  &=& \cot \theta_i \tan \theta_0 =\cot^{i} \theta_0 =\cot \theta_{i-1}.
\end{eqnarray*}
Thus, if we measure 0, we obtain the state $\ket{H_{i+1}}$ and if we measure 1, we obtain the state $\ket{H_{i-1}}$. The probability of measuring 0 is given by
\begin{eqnarray*}
  p_{0,i} &=& \cos^2 \theta_i \cos^2 \theta_0 +\sin^2 \theta_i \sin^2 \theta_0.
\end{eqnarray*}

\begin{figure}
  \includegraphics[width=8cm]{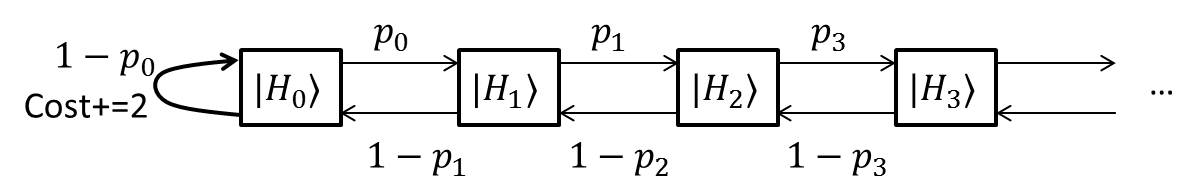}\\
  \caption{Process of obtaining other non-stabilizer states from initial $\ket H$ states. A copy of $\ket{H_i}$ and $\ket{H_0}$ probabilistically yield a copy of $\ket{H_{i-1}}$ or $\ket{H_{i+1}}$ using the circuit of Fig.~\ref{fig:2QbLadderCircs}. Each step along the ladder costs one copy of $\ket{H_0}$, except the first one which costs two.}\label{fig:RandomWalk}
\end{figure}

Note that $\frac{3}{4}\leq p_{0,i} <\cos^2 \frac\pi8=0.853\ldots$.
We can view this recursive process as a semi-infinite random walk with biased non-homogeneous probabilities, as Fig.~\ref{fig:RandomWalk} illustrates.
Every time a step is taken along this ``ladder" of states, one copy of $\ket H$ is consumed, except at the first node of the ladder (the production of state $\ket{H_0}$) when we require two copies of $\ket H$: if the outcome 1 is measured at the first node, we discard the output and start with two new copies of $\ket H$.

\begin{figure}
  \includegraphics[width=6cm]{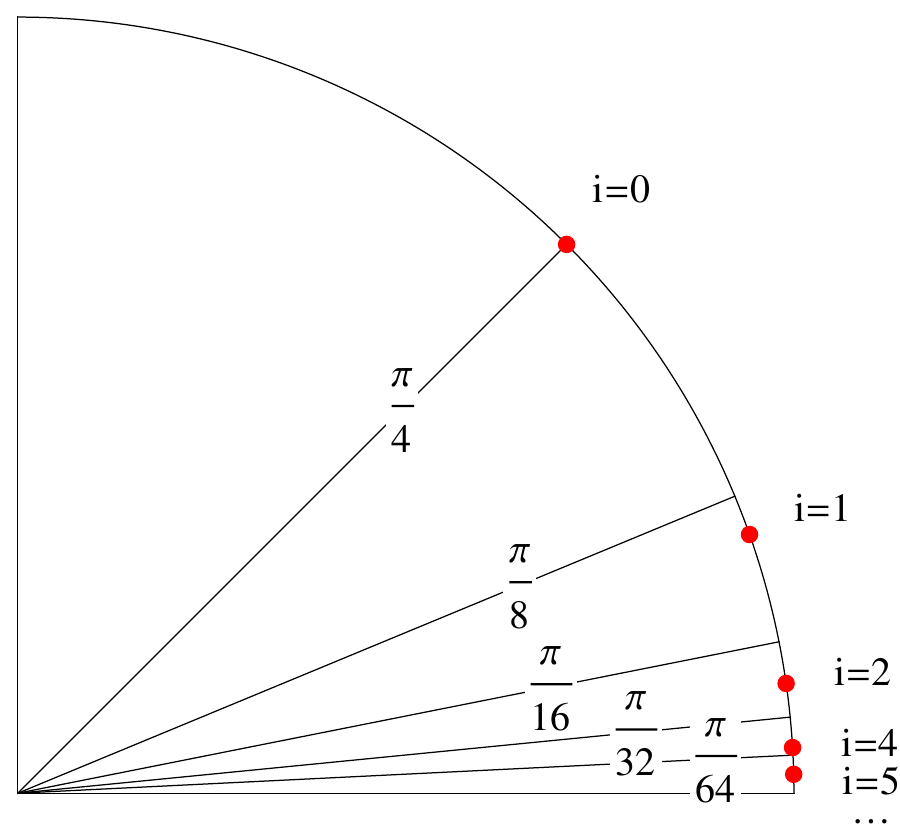}\\
  \caption{Red dots: Rotation by angle $2\theta_i$ implementable using an $\ket{H_i}=\cos\theta_i\ket 0 + \sin\theta_i\ket 1$ state.}\label{fig:AnglesCircle}
\end{figure}

Table \ref{tab:rots} lists the rotations obtained from the first few $i$ recursions, using the $\ket{H_i}$ states, and Figs.~\ref{fig:AnglesCircle}, \ref{fig:StateLadder} illustrate.
Note that there is a factor of two difference between the angle $\theta_i$ involved in the description of the state and the rotation applied, e.g., the $\ket{H_0}$ state is over $\theta_0=\frac\pi8$, and can be used to implement a $\frac\pi4$ rotation. Also, as $0<\theta_i<\frac{\pi}{4}$ $(\forall i)$, the discontinuity of the cotangent is never a problem.

\begin{table}
  \centering
  \begin{tabular}{|c|l||c|l|}
    \hline
    $i$ & $2\theta_i$ & $i$ & $2\theta_i$\\
    \hline\hline
    0 & $7.853\times10^{-1}$ & 9 & $2.974\times10^{-4}$\\
    1 & $3.398\times10^{-1}$ & 10 & $1.232\times10^{-4}$\\
    2 & $1.419\times10^{-1}$ & 11 & $5.102\times10^{-5}$\\
    3 & $5.886\times10^{-2}$ & 12 & $2.113\times10^{-5}$\\
    4 & $2.439\times10^{-2}$ & 13 & $8.753\times10^{-6}$\\
    5 & $1.010\times10^{-2}$ & 14 & $3.626\times10^{-6}$\\
    6 & $4.184\times10^{-3}$ & 15 & $1.502\times10^{-6}$\\
    7 & $1.733\times10^{-3}$ & 16 & $6.221\times10^{-7}$\\
    8 & $7.179\times10^{-4}$ & 17 & $\dots$\\
    \hline
  \end{tabular}
  \caption{Rotation by angle $2\theta_i$ implementable using an $\ket{H_i}=\cos\theta_i\ket 0 + \sin\theta_i\ket 1$ state.}\label{tab:rots}
\end{table}

\section{Numerical study of random $Z$-rotations}
\label{sec:Cost}

In this section, we numerically study the cost of implementing single-qubit rotations.
Although the circuit in Fig.~\ref{fig:RotCircs} randomly applies $\theta$ or $-\theta$, we show that our protocol still results in an efficient application of the desired $Z$-rotation.

Assume that we have the ideal case, where $\theta_i=2\theta_{i+1}$.
In this hypothetical scenario, if we try to apply some rotation using $\ket{H_i}$ and it fails,
then we can correct the gate by applying a rotation using $\ket{H_{i-1}}$.
If this gate also fails, then we follow with $\ket{H_{i-2}}$, and so on.
There are two crucial facts to point out.
First, the probability of failing $n$ times in a row scales as $1/2^n$, i.e., it decays exponentially with $n$, such that the expected number of iterations is well-behaved.
Second, and very importantly, if a $T$ gate fails (using a $\ket{H_0}$ state),
we can always correct deterministically by applying the phase gate $S$ (which is a $Z$-rotation by $\rme^{i\pi/2}$).

\begin{figure}
  \includegraphics[width=6 cm]{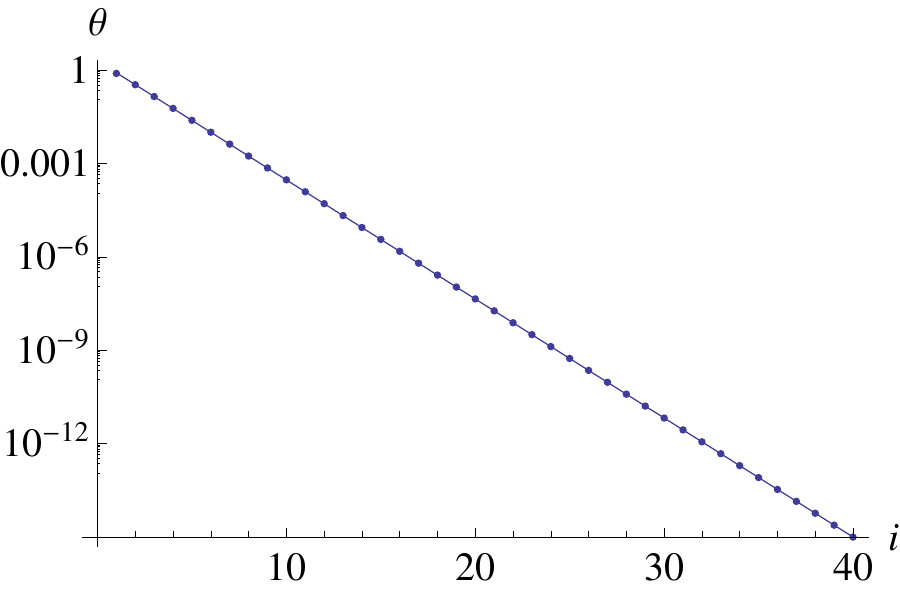}\\
  \caption{Dots: States obtainable by recursively using the circuit of Fig.~\ref{fig:2QbLadderCircs} and only $\ket H$ states as initial input. Full line: exponential decay fit, $\theta_i\sim2.41^{-0.881i}$.}\label{fig:StateLadder}
\end{figure}

Unfortunately, $\theta_i\neq2\theta_{i+1}$; however, this assumption is not too far from the truth (see Fig.~\ref{fig:StateLadder}).
Since we only require that a gate be approximated to a given precision, we can actually apply any $Z$-rotation rapidly and with good accuracy.
First, apply the $\ket{H_i}$-rotation that gets you closest to your target angle and then recurse.
The next gate to apply will depend on whether the current gate succeeded or not.
Due to the property discussed in the previous paragraph, we will rapidly converge towards our target angle.

We now present simulation results of obtaining the Z-rotation $Z(\phi)$, where $\phi$ is chosen randomly, in order to characterize the efficiency of the protocol proposed in previous sections.
The simulation proceeds as follows:
\begin{enumerate}
    \item Set desired accuracy $\epsilon$.
    \item Randomly pick a target rotation angle $0<\phi<2\pi$.
    \item Find the state $\ket{H_i}$ such that $2\theta_i$ is close to $\phi$.
    \item Simulate an instance of the ladder to obtain that state and add its cost to the offline cost.
    \item Apply a rotation using the $\ket{H_i}$ state and the circuit of Fig.~\ref{fig:RotCircs} and add one to the online cost.
    \item Recurse on steps 3 through 5 until the desired accuracy is reached.
\end{enumerate}

We define the accuracy of the applied rotation $V$ compared to the target rotation $U$ as
\begin{eqnarray*}
    \max_{\ket\psi}D(U\ket{\psi}\bra{\psi}U^\dagger,V\ket{\psi}\bra{\psi}V^\dagger),
\end{eqnarray*}
where
\begin{eqnarray*}
    D(\rho,\sigma) &=& \frac12 \mathrm{tr}\left(\sqrt{(\rho-\sigma)^\dagger(\rho-\sigma)}\right)
\end{eqnarray*}
is the trace
 distance between states $\rho$ and $\sigma$.
If $U$ and $V$ are rotations about the same axis, one can show that for small angles of rotation, which will always be our case, this reduces to the difference of rotation angles, $\epsilon=\Delta\phi$.

In \cite{BS12}, the distance measure used is
\begin{eqnarray*}
    D(U,V) &=& \sqrt{\frac{2-|\mathrm{tr}(UV^\dagger)|}{2}}.
\end{eqnarray*}
In the case of rotations about the same axis, it can be reduced to $\sqrt{1-|\cos(\Delta\phi)|}\approx\Delta\phi/\sqrt2$ for small $\Delta\phi$.
This conversion between the two measures is important since we later compare performance.

To compare the resource cost of our protocol to Solovay-Kitaev decomposition, we define an online and offline cost to apply a unitary gate.
The \emph{online cost}, $C_{\textrm{on}}$, is the expected number of non-Clifford gates, or $\ket{H_i}$ states, required to implement the unitary gate on a qubit.
The \emph{offline cost}, $C_{\textrm{off}}$, is the total number of distilled $\ket H$ states required to obtain all of the intermediate $\ket{H_i}$ states used to perform the given unitary operation, that is, the sum of the $\ket H$ states used for each ladder process.
In our resource costs, we do not include the initial cost to distill $\ket H$ states.
For Solovay-Kitaev decomposition, the offline cost is always 0 and the online cost is the total number of $T$ and $T^\dagger$ gates in the decomposition.

We ran the simulation for target accuracies ranging between $10^{-12}<\epsilon<10^{-4}$, each time considering a new random angle to produce a sample of $\sim1.8\times10^4$ instances of this protocol. Just like in the case of Solovay-Kitaev decomposition, we suppose that
\begin{eqnarray*}
  C_{\textrm{on}} &\sim& \ln^c(\frac1\epsilon),\\
  C_{\textrm{off}} &\sim& \ln^{c'}(\frac1\epsilon),
\end{eqnarray*}
where $C_{\textrm{on}}$ and $C_{\textrm{off}}$ are the online and offline costs, respectively, such that
\begin{eqnarray*}
  \ln C_{\textrm{on}} &\sim& c\ln\ln(\frac1\epsilon),\\
  \ln C_{\textrm{off}} &\sim& c'\ln\ln(\frac1\epsilon).
\end{eqnarray*}

\begin{figure}
    \centering
    \subfigure[Fit: $\ln(C_\textrm{on})=-0.49 + 1.29\ln(\ln(1/\epsilon))$.]{
        \includegraphics[width=6 cm]{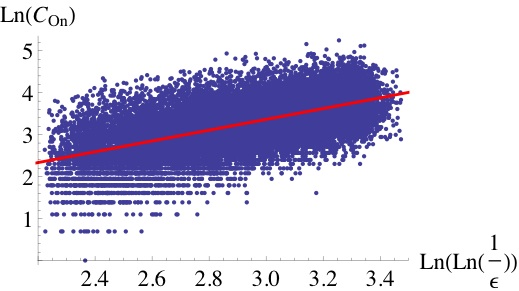}
        \label{fig:ScalingOnline}
    }
    \subfigure[Fit: $\ln(C_\textrm{Off})=-0.72 + 2.27\ln(\ln(1/\epsilon))$.]{
        \includegraphics[width=6 cm]{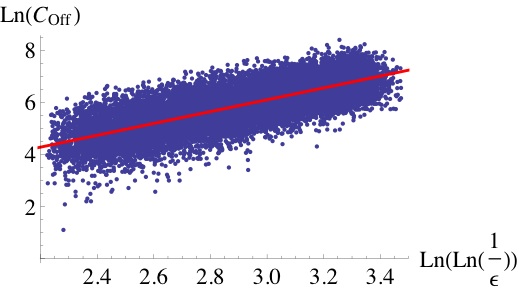}
        \label{fig:ScalingOffline}
    }
  \caption{Target accuracies are chosen such that $10^{-12}<\epsilon<10^{-4}$ and the sample size is $\sim 1.8\times10^4$. The clouds of points are used to fit the data according to a linear fit. We obtain $c \sim 1.29$ and $c' \sim 2.27$.}\label{fig:Scaling}
\end{figure}

The results are given in Fig.~\ref{fig:Scaling}. Fits are also presented from which we deduce that $c \sim 1.29$ and $c' \sim 2.27$ for our protocol.

As discussed in Section \ref{sec:SK}, both of these scalings represent a significant improvement over the best implementation, to our knowledge, of Solovay-Kitaev decomposition \cite{BS12}, which was itself a significant improvement over the previous implementation of \cite{DN05}.

Note that one can implement any single-qubit unitary $U$ using three rotations around the $X$- and $Z$-axes \cite{NC00}:
\begin{eqnarray*}
  U &\propto&  X(\alpha)Z(\beta)X(\gamma),
\end{eqnarray*}
for some angles $\alpha,\beta,\gamma$.
We have explicitly shown simulaton results for $Z$-rotations, however $X$-rotations can be obtained at the same cost using the $X$-rotation circuit given in Fig.~\ref{fig:RotCircs}.
Thus we can use our protocol to produce each of the three rotations, and produce any desired single-qubit unitary operation.

\subsection{Other states}

To further reduce the resource costs and their respective scalings, we show that we can use different Clifford circuits to produce new non-Clifford states that can be used as initial resources for the previously presented protocols.
We first introduce three new states $\ket{\psi^{0,1,2}_0}$ and discuss how to combine them into the described protocol.

\begin{figure}
\[
\Qcircuit @C=1em @R=.5em {
\lstick{\ket{H_0}} & \gate{H}  & \qw         & \gate{X} & \qw  & \qw & \ctrl{3}  & \meter & \rstick{0} \cw \\
\lstick{\ket{H_0}} & \qw       & \ctrl{1}    & \qw   &    \ctrl{2}  & \gate{H} & \qw & \meter & \rstick{0} \cw \\
\lstick{\ket{H_0}} & \gate{H}  & \gate{X}    & \ctrl{-2} & \qw &  \ctrl{1} & \qw & \gate{H}  & \rstick{H\ket{\psi^0}} \qw \\
\lstick{\ket{H_0}} & \qw       & \qw         & \qw      &  \gate{X} & \gate{Z} & \gate{X} & \meter & \rstick{0} \cw \\
}\]
\caption{Circuit to produce $\ket{\psi^0_0}$ states. The probability of success is $0.320$ and every trial consumes four copies of $\ket H$ such that the average cost is $12.50$ to produce a copy of $\ket{\psi^0_0}$.}\label{fig:Psi0Circ}
\end{figure}
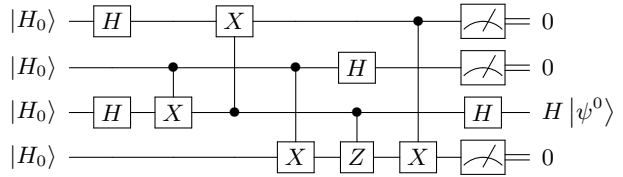

Consider the circuit of Fig.~\ref{fig:Psi0Circ}.
It is a Clifford circuit to which we input four copies of $\ket H$.
The measurement outcome $000$ occurs with probability $3(2+\sqrt2)/32\approx0.320$, otherwise the output is discarded.
If the measurement yields result 000, then the produced state is
\begin{eqnarray*}
  \ket{\psi^0_0} &=& \cos\phi^0_0 \ket0+\sin\phi^0_0\ket1,\\
  \phi^0_0 &=& \frac\pi2-\cot^{-1}\left(\frac{2+3\sqrt2}{6+5\sqrt2}\right)\approx 0.446.
\end{eqnarray*}

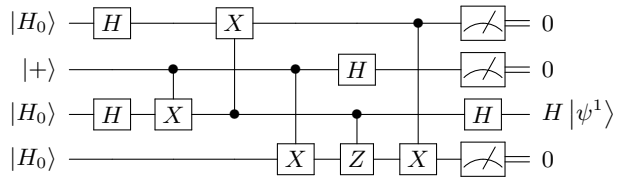
\begin{figure}
\[
\Qcircuit @C=1em @R=.5em {
\lstick{\ket{H_0}} & \gate{H}  & \qw         & \gate{X} & \qw  & \qw & \ctrl{3}  & \meter & \rstick{0} \cw \\
\lstick{\ket{+}} & \qw       & \ctrl{1}    & \qw   &    \ctrl{2}  & \gate{H} & \qw & \meter & \rstick{0} \cw \\
\lstick{\ket{H_0}} & \gate{H}  & \gate{X}    & \ctrl{-2} & \qw &  \ctrl{1} & \qw & \gate{H}  & \rstick{H\ket{\psi^1}} \qw \\
\lstick{\ket{H_0}} & \qw       & \qw         & \qw      &  \gate{X} & \gate{Z} & \gate{X} & \meter & \rstick{0} \cw \\
}\]
 \caption{Circuit to produce $\ket{\psi^1_0}$ states. The probability of success is $0.232$ and every trial consumes four copies of $\ket H$ such that the average cost is $12.95$ to produce a copy of $\ket{\psi^1_0}$.}\label{fig:Psi1Circ}
\end{figure}

Since the probability of success is $0.320$ and that every trial consumes four copies of $\ket H$, the average cost to produce $\ket{\psi^0_0}$ is $12.50$ $\ket H$ states. This circuit was designed to measure the stabilizer code presented in Table \ref{tab:Psi0Stab}.
Another interesting state can be obtained from the same circuit, substituting one of the input states by a $\ket +$ state as is illustrated by Fig.~\ref{fig:Psi1Circ}.
The measurement outcome 000 is obtained with probability $(6+\sqrt2)/32\approx0.232$. The corresponding output state is
\begin{eqnarray*}
  \ket{\psi^1_0} &=& \cos\phi^1_0 \ket0+\sin\phi^1_0 \ket1,\\
  \phi^1_0 &=& \frac\pi2-\cot^{-1}\left(\frac{2\sqrt2}{3+\sqrt2}\right)\approx 0.570.
\end{eqnarray*}

\begin{figure}
\[
\Qcircuit @C=1em @R=.5em {
\lstick{\ket{H_0}} & \qw         & \ctrl{3} & \ctrl{2}  & \gate{H}  & \meter & \rstick{0} \cw \\
\lstick{\ket{H_0}} & \gate{X}    & \qw      & \qw       & \gate{X} & \gate{H} & \rstick{\ket{\psi^2}} \qw \\
\lstick{\ket{H_0}} & \ctrl{-1}    & \qw     & \gate{X}  & \qw & \meter & \rstick{0} \cw \\
\lstick{\ket{H_0}} & \qw       & \gate{X}   & \qw      &  \ctrl{-2} & \meter & \rstick{0} \cw \\
}\]
  \caption{Circuit to produce $\ket{\psi^2_0}$ states. The probability of success is $0.344$ and every trial consumes four copies of $\ket H$ such that the average cost is $11.64$ to produce a copy of $\ket{\psi^2_0}$.}\label{fig:Psi2Circ}
\end{figure}
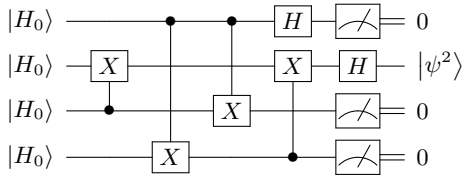

Since the probability of success is $0.232$ and every trial consumes three copies of $\ket H$, the average cost to produce $\ket{\psi^1_0}$ is $12.95$ $\ket H$ states.
Fig.~\ref{fig:Psi2Circ} presents another useful circuit. The measurement outcome 000 is obtained with probability $11/32\approx0.344$. The corresponding output state is
\begin{eqnarray*}
  \ket{\psi^2_0} &=& \cos\phi^2_0 \ket0+\sin\phi^2_0 \ket1,\\
  \phi^2_0 &=& \frac\pi2-\cot^{-1}\left(\frac{7}{6\sqrt2}\right)\approx 0.690.
\end{eqnarray*}

The probability of success is $0.344$ and every trial consumes four copies of $\ket H$ such that the average cost to produce $\ket{\psi^2_0}$ is $11.64$ $\ket H$ states.
Table \ref{tab:Psi2Stab} presents the stabilizer code in terms of its generators $S$ that are decoded by the circuit.

%

\begin{table}
  \centering
  \begin{tabular}{|c|c c c c c|}
    \hline
    $S$ & $\pm$ & 0 & 1 & 2 & 3\\
    \hline\hline
    $s_0$ & + & X & Z & X & .\\
    $s_1$ & + & . & X & Z & X\\
    $s_2$ & + & X & . & X & Z\\
    $\overline{Z}$ & + & Z & Z & Z & Z\\
    \hline
  \end{tabular}
  \caption{The stabilizer code decoded by the circuit of Fig.~\ref{fig:Psi0Circ}.}\label{tab:Psi0Stab}
\end{table}

\begin{table}
  \centering
  \begin{tabular}{|c|c c c c c|}
    \hline
    $S$ & $\pm$ & 0 & 1 & 2 & 3\\
    \hline\hline
    $s_0$ & + & X & X & X & X\\
    $s_1$ & + & Z & . & Z & .\\
    $s_2$ & + & Z & . & . & Z\\
    $\overline{Z}$ & + & Z & Z & Z & Z\\
    \hline
  \end{tabular}
  \caption{The stabilizer code decoded by the circuit of Fig.~\ref{fig:Psi2Circ}.}\label{tab:Psi2Stab}
\end{table}

We will use these states as input states to the circuit given in Fig.~\ref{fig:2QbLadderCircs}, where one of these states is used in place of the $\ket{H_0}$ input state. We start with a copy of $\ket{\psi^i_0}$ and a copy of $\ket{H_0}$. If measurement outcome 1 is obtained, the state is discarded. Otherwise, we obtain
\begin{eqnarray*}
  \ket{\psi^i_1} &=& \cos\phi^i_1 \ket0+\sin\phi^i_1 \ket1,\\
  \cot\phi^i_1 &=& \cot\phi^i_0 \cot \theta_0.
\end{eqnarray*}
Similarly to the $\ket{H_i}$ states, we define
\begin{eqnarray*}
  \ket{\psi^j_i} &=& \cos\phi^j_i \ket0+\sin\phi^j_i \ket1,\\
  \cot\phi^j_i &=& \cot\phi^j_0 \cot^i \theta_0.
\end{eqnarray*}
If we input a copy of $\ket{\psi^j_i}$ and a copy of $\ket{H_0}$, we obtain
\begin{eqnarray*}
  \ket{H_0}\ket{\psi^j_i} &\xrightarrow{\Lambda{(X)}}& \cos \theta_0 \cos \phi^j_i \ket{00}+ \sin \theta_0 \sin \phi^j_i \ket{01}\\
   & & + \sin \theta_0 \cos \phi^j_i \ket{10}+ \cos \theta_0 \sin \phi^j_i \ket{11}.
\end{eqnarray*}
such that the output state obtained is, depending on measurement outcome,
\begin{eqnarray*}
 &\xrightarrow{m=0}&  \ket{\psi^j_{i+1}}\\
 &\xrightarrow{m=1}&  \ket{\psi^j_{i-1}}.
\end{eqnarray*}

\begin{figure}
  \includegraphics[width=6 cm]{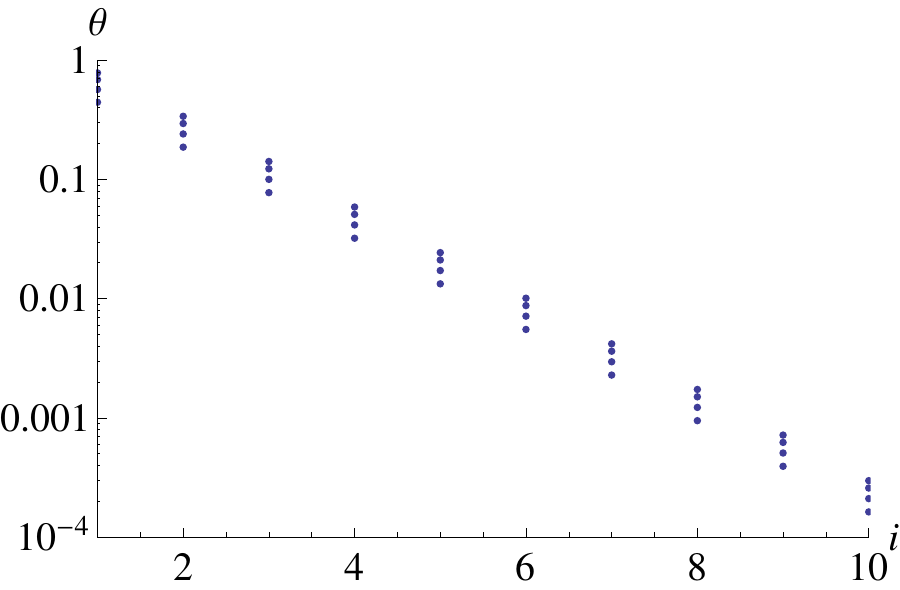}\\
  \caption{Dots: States obtainable by recursively using the circuit of Fig.~\ref{fig:2QbLadderCircs} with initial resource states $\ket H$, $\ket{\psi^0}$, $\ket{\psi^1}$ and $\ket{\psi^2}$.}\label{fig:StateLadderMulti}
\end{figure}

\begin{figure}
  \includegraphics[width=6cm]{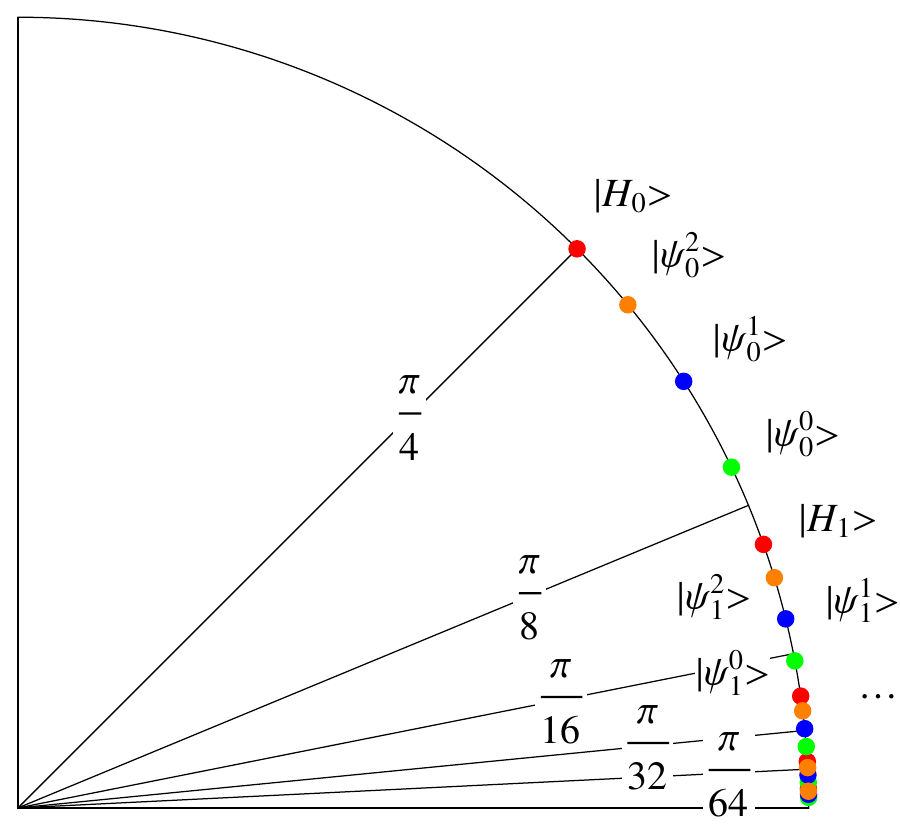}\\
  \caption{Dots: Rotations implementable using $\ket{H_i}$, $\ket{\psi^0_i}$, $\ket{\psi^1_i}$, $\ket{\psi^2_i}$ states.}\label{fig:AnglesCircleMulti}
\end{figure}

\begin{figure}
    \centering
    \subfigure[Fit: $\ln(C'_\textrm{on})=-0.78 + 1.12\ln(\ln(1/\epsilon))$.]{
        \includegraphics[width=6 cm]{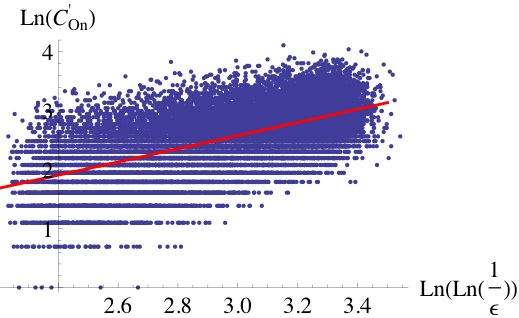}
        \label{fig:ScalingOnlineMulti}
    }
    \subfigure[Fit: $\ln(C'_\textrm{Off})=0.54 + 1.75\ln(\ln(1/\epsilon))$.]{
        \includegraphics[width=6 cm]{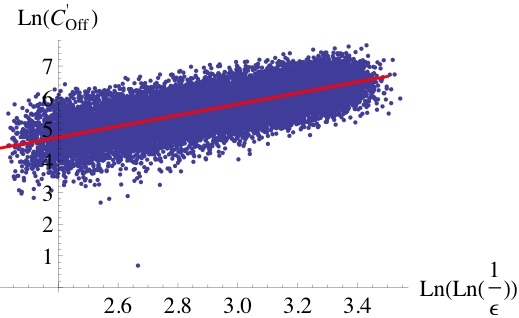}
        \label{fig:ScalingOfflineMulti}
    }
  \caption{Target accuracies are chosen such that $10^{-12}<\epsilon<10^{-4}$ and the sample size is $\sim 1.8\times10^4$. The clouds of points are used to fit the data according to a linear fit. We obtain $c \sim 1.12$ and $c' \sim 1.75$, improving on the results of Fig.\ref{fig:Scaling}.}\label{fig:ScalingMulti}
\end{figure}

New ``ladders" of states can be obtained using the $\ket{\psi^{0,1,2}_0}$ states as inputs in place of the $\ket{H_0}$ states. Fig.~\ref{fig:StateLadderMulti} shows the four ladders. Table \ref{tab:rotsMulti} lists the rotations obtained from the first few $i$ recursions and Fig.\ref{fig:AnglesCircleMulti} illustrates.
We see that the set of possible rotations is more dense.
We reproduced the numerical experiment of Section \ref{sec:Cost} with basic offline costs of 12.50, 12.95 and 11.64 for $\ket{\psi^0_0}$, $\ket{\psi^1_0}$, and $\ket{\psi^2_0}$, respectively.
The results are presented in Fig.~\ref{fig:ScalingMulti}.
Since the set of states is denser, we expected improved scalings for both the online and offline costs. This is indeed the case, we find $c \sim 1.12$ and $c' \sim 1.75$.

\begin{figure}
  \includegraphics[width=6 cm]{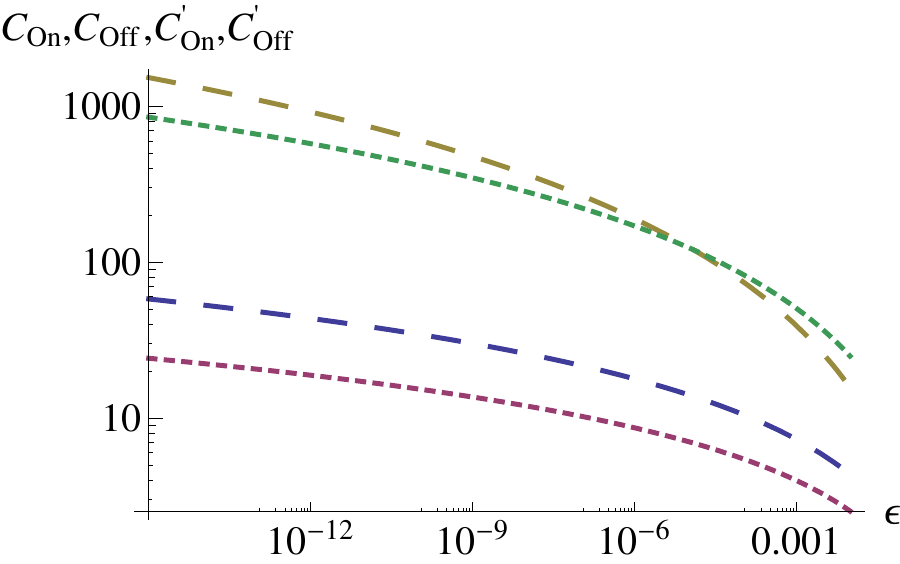}\\
  \caption{Dashed line: Offline (top) and online (bottom) costs for the scheme using only $\ket H$ states. Dotted lines: Offline (top) and online (bottom) costs for the scheme using $\{\ket H,\ket{\psi^0_0},\ket{\psi^1_0},\ket{\psi^2_0}\}$ states. The two top curves cross at $\epsilon\approx1.28\times10^{-5}$.}\label{fig:ComparisionHMulti}
\end{figure}

However, the basic offline costs of our new states $\ket{\psi^i_0}$ are significantly higher;
for precision $\sim10^{-4}$, even though the online cost is smaller using the new states, the offline cost is still smaller if we restrict ourselves to the simpler scheme using only $\ket H$ states.
For the protocol using the new input states to reduce both the online and offline costs, we need to consider precisions smaller then $\epsilon\approx1.28\times10^{-5}$, see Fig.~\ref{fig:ComparisionHMulti}.

\begin{table}
  \centering
  \begin{tabular}{|c||l|l|l|l|}
    \hline
    $i$ & $2\theta_i$ & $2\phi^0_i$ & $2\phi^1_i$ & $2\phi^2_i$\\
    \hline\hline
    0 & $7.853\times10^{-1}$ & $4.456\times10^{-1}$ & $5.698\times10^{-1}$ & $6.898\times10^{-1}$\\
    1 & $3.398\times10^{-1}$ & $1.871\times10^{-1}$ & $2.415\times10^{-1}$ & $2.954\times10^{-1}$\\
    2 & $1.419\times10^{-1}$ & $7.770\times10^{-2}$ & $1.004\times10^{-1}$ & $1.231\times10^{-1}$\\
    3 & $5.886\times10^{-2}$ & $3.220\times10^{-2}$ & $4.162\times10^{-2}$ & $5.105\times10^{-2}$\\
    4 & $2.439\times10^{-2}$ & $1.334\times10^{-2}$ & $1.724\times10^{-2}$ & $2.115\times10^{-2}$\\
    5 & $1.010\times10^{-2}$ & $5.525\times10^{-3}$ & $7.142\times10^{-3}$ & $8.761\times10^{-3}$\\
    6 & $4.184\times10^{-3}$ & $2.288\times10^{-3}$ & $2.959\times10^{-3}$ & $3.629\times10^{-3}$\\
    7 & $1.733\times10^{-3}$ & $9.479\times10^{-4}$ & $1.225\times10^{-3}$ & $1.503\times10^{-3}$\\
    8 & $7.179\times10^{-4}$ & $3.926\times10^{-4}$ & $5.076\times10^{-4}$ & $6.226\times10^{-4}$\\
    \hline
  \end{tabular}
  \caption{Rotations implementable using $\ket{H_i},\ket{\psi^0_i},\ket{\psi^1_i},\ket{\psi^2_i}$ states.}\label{tab:rotsMulti}
\end{table}

The reason why we do not consider other measurement outcomes in the circuits considered in this section is that in general, potential errors on the $\ket H$ states are amplified by the circuit.
Output states in these cases might still prove useful, but a careful analysis of the evolution of errors must be conducted.

\subsection{Minimizing the online cost}

In this section, we aim to minimize the online cost, at the price of potentially increasing the offline cost.

The protocol up to this point can be summarized as follows.
Suppose one wants to implement a $Z$-rotation of an arbitrary angle $\phi$ on a logical state $\ket\psi$. One has to implement a sequence of $j$ rotations $\{Z(2\theta_{i_j})\}$ on $\ket\psi$ using the sequence of states $\{\ket{H_{i_j}}\}$, such that $Z(\phi)\approx\prod_j Z(2\theta_{i_j})$. The online cost is given by $\left|\{\ket{H_{i_j}}\}\right|$.

Consider instead the following protocol to implement the same rotation by angle $\phi$. Prepare \emph{offline} the state $\ket{Z(\phi)}$ from a copy of $\ket0$. To achieve this, use the protocol described in the previous paragraph to rotate $\ket0$ to $\ket{Z(\phi)}$. Note that you can implement it offline because you are applying rotations on an ancilla state. Then, use $\ket{Z(\phi)}$ \emph{online} to apply the desired rotation. With probability $\frac12$, the rotation $Z(\phi)$ is applied and the online cost is 1. If it fails, correct for it by preparing \emph{offline} the state $\ket{Z(2\phi)}$. Again, with probability $\frac12$, the overall rotation $Z(\phi)$ is applied and the online cost is 2. If it fails, prepare \emph{offline} $\ket{Z(4\phi)}$, and so on. The probability that a number $n$ of iterations is required before success decreases exponentially with $n$. This process is a negative binomial of parameter $p=\frac12$ and the expected number of online rotations before success goes as $\sim\frac1p=2$.

\begin{figure}
    \centering
    \subfigure[Fit: $\langle C''_\textrm{on}\rangle=1.99$.]{
        \includegraphics[width=6 cm]{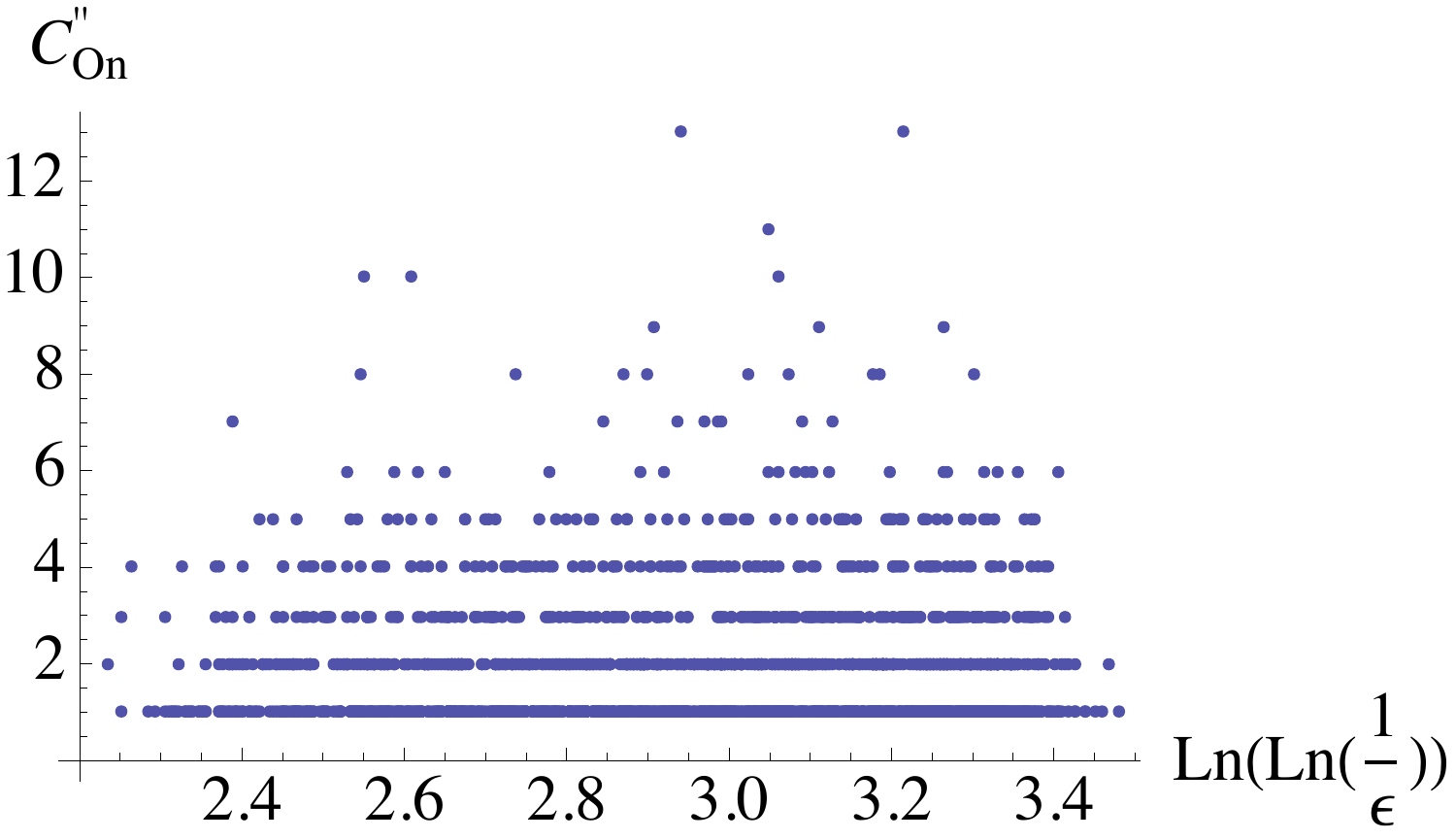}
        \label{fig:ScalingOnlineMinOnCost}
    }
    \subfigure[Fit: $\ln(C''_\textrm{Off})=1.13 + 1.75\ln(\ln(1/\epsilon))$.]{
        \includegraphics[width=6 cm]{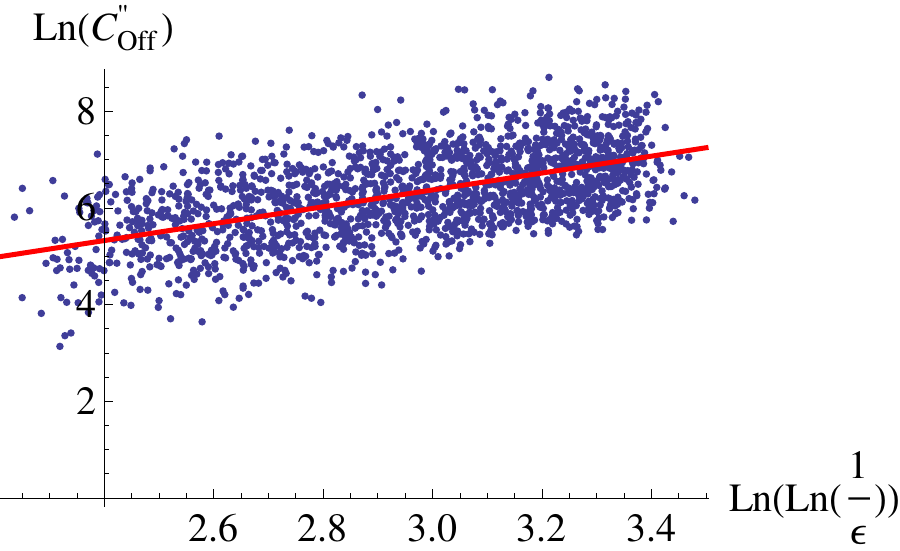}
        \label{fig:ScalingOfflineMinOnCost}
    }
  \caption{Target accuracies are chosen such that $10^{-12}<\epsilon<10^{-4}$ and the sample size is $\sim 1.8\times10^3$. (a) The cloud of points is averaged to get the expected number of online rotations. (b) The cloud of point is used to perform a linear fit.}\label{fig:ScalingMinOnCost}
\end{figure}

We numerically simulated this process for various random angles, $0<\phi<2\pi$, and accuracies, $10^{-12}<\epsilon<10^{-4}$. As Fig.~\ref{fig:ScalingMinOnCost} illustrates, we note that there is no significant change in the online cost for different values of $\epsilon$.
Moreover, the expected number of online rotations to apply is roughly two, the result one would expect in this situation.
The scaling of the offline cost is the same as that of the previous scheme, as expected.
Also, even though the scaling is the same, the actual values of the offline cost are bigger. The shift is $1.13-0.64=0.59$, see Figs. \ref{fig:ScalingOfflineMulti} and \ref{fig:ScalingOfflineMinOnCost}, which corresponds to a factor of $\rme^{0.59}\approx1.80$.
One would expect a slightly bigger shift of the offline cost by $\ln 2$ since we repeat the scheme twice on average.
This suggests that there might exist some favorable correlations between the expected offline cost of an angle $\theta$ and of twice that angle $2\theta$.

\section{Erroneous states}

In this section, we determine the effect of errors on our resource $\ket H$ states, and their effect on the produced $\ket{H_i}$ states.
A priori, the errors might be amplified by the two-qubit circuit of Fig.~\ref{fig:2QbLadderCircs}, however we show this is not the case.

The probabilistic and non-homogeneous nature of the presented protocol is not well suited for an analytical study of the evolution of errors on $\ket{H_i}$ states.
Instead, we rely on a numerical study for three different types of errors. We use the trace distance on states $\rho$ and $\sigma$,
\begin{eqnarray*}
    D(\rho,\sigma) &=& \frac12 \mathrm{tr}(\sqrt{(\rho-\sigma)^\dagger(\rho-\sigma)}),
\end{eqnarray*}
to measure the accuracy of the imperfect $\ket{H_i}$ states.
We assume Clifford operations are perfect and that errors can only occur on the $\ket H$ states.

We consider three types of erroneous states.
First, we assume that the mixed state, $\rho^a_0$, is perfectly along the line joining the center of the Bloch sphere and the the perfect state, i.e.,
\begin{eqnarray*}
    \rho^a_0(p) &=& (1-p)|H_0\rangle\langle H_0|+p|-H_0\rangle\langle -H_0|,
\end{eqnarray*}
where $\ket{-H_0}=\sin\frac\pi8 \ket0-\cos\frac\pi8 \ket1$ is the state orthogonal to $\ket{H_0}$.
We denote the imperfect version of $\ket{H_i}$ obtained from $\rho^a_0$ states as $\rho^a_i$.
If Clifford operations are perfect, we can always bring any mixed state into this form using twirling \cite{MEK05}.
However, for the protocol to be of practical interest, we require it to remain stable under the two following types of errors, where we assume that the state is pure, but that the rotation is slightly off of the desired axis by $\delta$:
\begin{eqnarray*}
    \rho^b_0(\delta) &=& \frac12\left(I+\sin\left(\frac\pi4+\delta\right)X+\cos\left(\frac\pi4+\delta\right)Z\right)\\
    \rho^c_0(\delta) &=& \frac12\left(I+\sin\frac\pi4 \cos\delta X+\sin\frac\pi4 \sin\delta Y+\cos\frac\pi4 Z\right).
\end{eqnarray*}

\begin{figure}
  \includegraphics[width=6 cm]{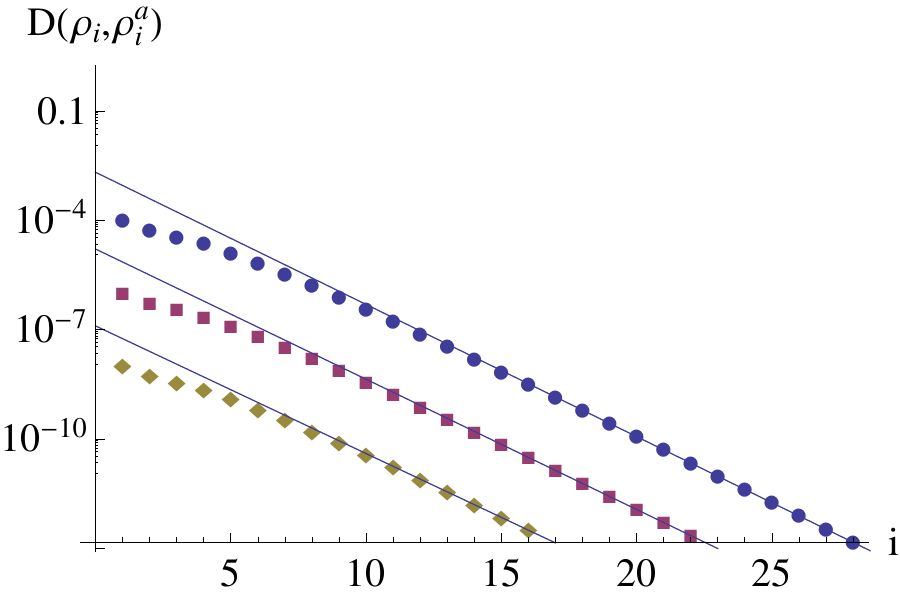}\\
  \caption{Evolution of the trace distance between imperfect $\rho^a_i$ and perfect $\ket{H_i}$ states. Circles: data for $p=10^{-4}$, $1\leq i\leq 28$. Squares: data for $p=10^{-6}$, $1\leq i\leq 22$. Diamonds: data for $p=10^{-8}$, $1\leq i\leq 16$. The full lines are exponential decay fits: $(2.08*10^{-3})\times2.31^{-i}$ using points $18\leq i\leq 28$, $(1.63*10^{-5})\times2.28^{-i}$ using points $18\leq i\leq 22$ and $(1.26*10^{-7})\times2.24^{-i}$ using points $13\leq i\leq 16$ for the circle, square and diamond data set, respectively. Sample size is $1000$. We conclude that if the initial resource state has desired accuracy, then this is also true of all derived resource states.}\label{fig:DistConva}
\end{figure}

\begin{figure}
    \centering
    \subfigure[Fits: $(1.17*10^{-3})\times2.31^{-i}$, $(1.03*10^{-5})\times2.29^{-i}$ and $(7.50*10^{-8})\times2.25^{-i}$]{
         \includegraphics[width=6 cm]{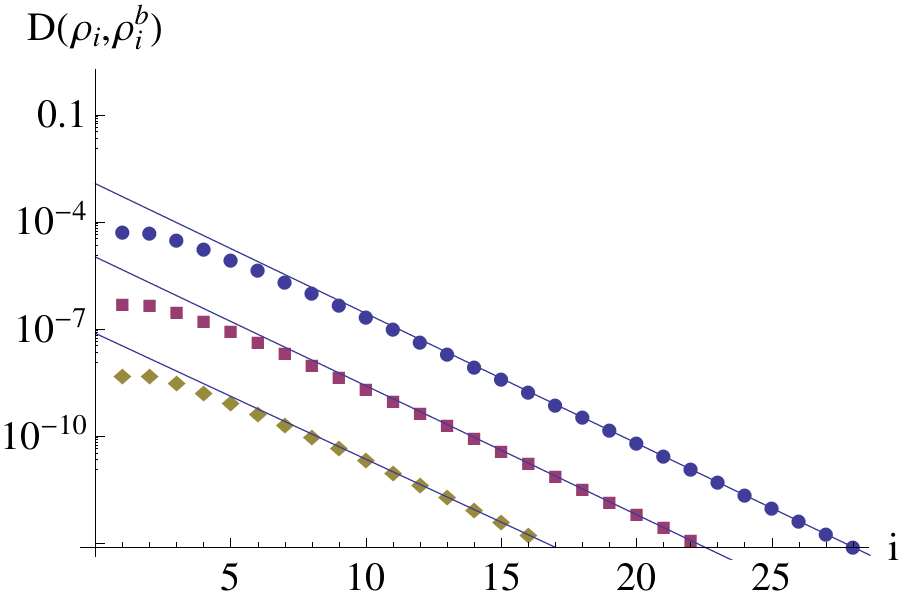}
        \label{fig:DistConvb}
    }
        \subfigure[Fits: $(8.28*10^{-4})\times2.31^{-i}$, $(7.32*10^{-6})\times2.30^{-i}$ and $(5.30*10^{-8})\times2.25^{-i}$]{
         \includegraphics[width=6 cm]{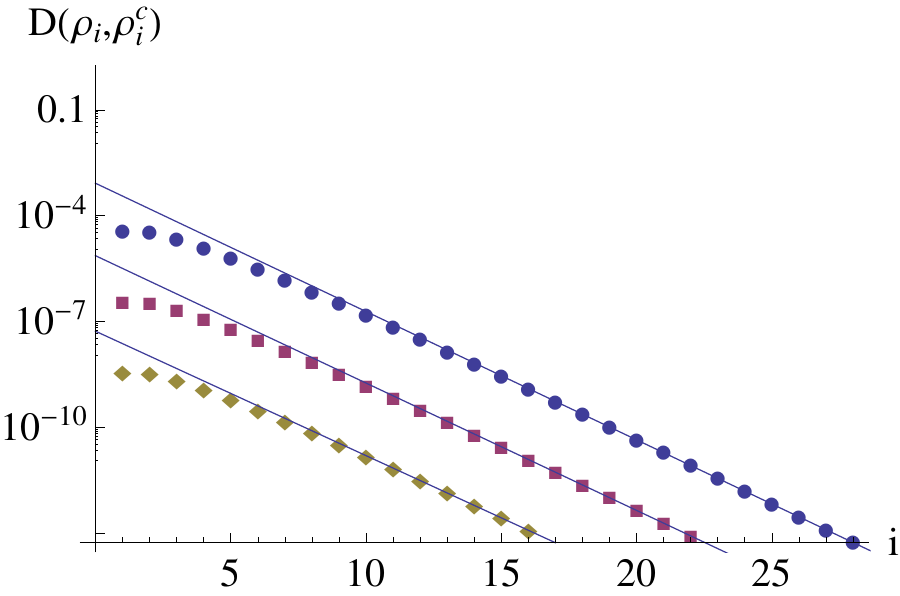}
        \label{fig:DistConvc}
    }
  \caption{Distances between the ideal $\ket{H_i}$ states and the imperfects states $\rho^b_i$ and $\rho^c_i$ respectively. Sample size is $1000$. $\epsilon=10^{-4},10^{-6},10^{-8}$ in both cases.}\label{fig:DistConvbc}
\end{figure}

We numerically generated pseudo-random instances of the scheme to produce $\ket{H_i}$ states for different values of $i$ and for different noise strengths. We considered 1000 instances for each of the three types of errors and noise strengths $10^{-4},10^{-6}$ and $10^{-8}$. Figures \ref{fig:DistConva} and \ref{fig:DistConvbc} show that the protocol actually reduces the amplitude of possible errors on the resource $\ket H$ states, such that if we start with $\ket H$ meeting our target accuracy, all the subsequent derived $\ket{H_i}$ states will also meet it. This even suggests that for bigger values of $i$, one could use noisier $\ket H$ states and still achieve the desired accuracy. This could make a dramatic difference if it enables one to reduce the number of distillation recursions necessary to prepare the $\ket H$ states.

We note a very similar behavior for the three types of errors. The exponential decay of the distance between erroneous and ideal states confirms that the errors are well behaved under the proposed protocol.
We note that the bases for the exponential decay of the errors are comparable, but smaller, than the basis for the exponential decay of the angle implemented. So, for a given error rate, there exits a point where the angle of rotation implemented by $\ket{H_i}$ for some $i$ is going to be comparable or smaller to the error on that angle. However, this is not a problem in practice since for, e.g., $\epsilon\sim10^{-4}$, we find $i=150$. For this value of $i$ the angle is $\theta_i\sim10^{-57}$.

\section{Comparison to Solovay-Kitaev decomposition}
\label{sec:SK}

In this section, we compare the performance of the Solovay-Kitaev decomposition of \cite{BS12} and that of the schemes presented in this article. In order to do this, we first consider different $Z$-rotations of angles $\pi/16$, $\pi/128$, and $\pi/1024$ and different accuracies $10^{-4}, 10^{-8}$ and $10^{-12}$.


\begin{figure}
  \subfigure[\newline$\ln(C^Z_\textrm{SK})=-4.88 + 4.41\ln(\ln(1/\epsilon))$\newline$\ln(C^Z_\textrm{On})=-0.49 + 1.29\ln(\ln(1/\epsilon))$\newline$\ln(C^Z_\textrm{Off})=-0.72 + 2.27\ln(\ln(1/\epsilon))$\newline]{
    \includegraphics[width=6 cm]{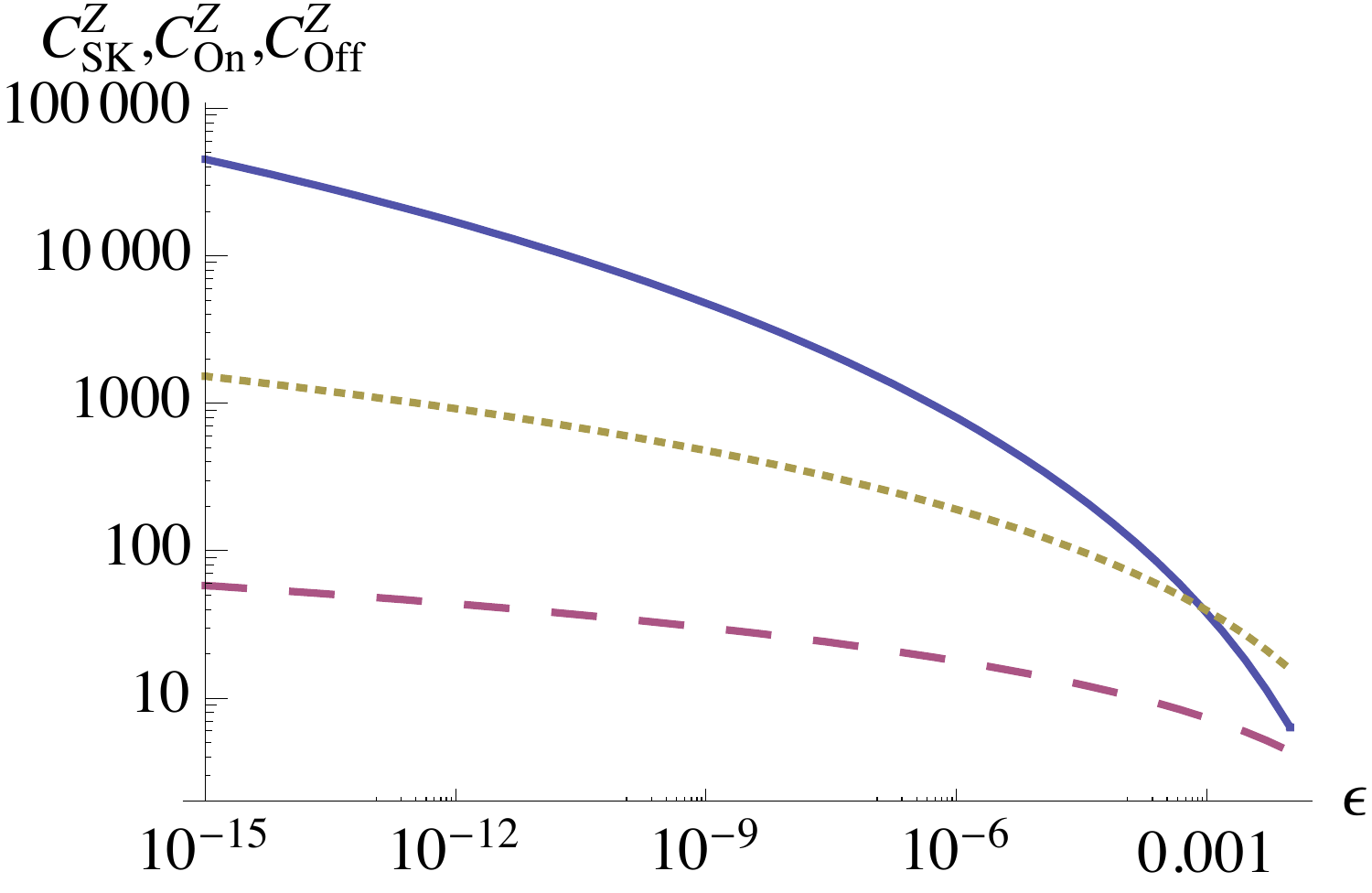}
    \label{fig:ComparisonSKZ}
  }
  \subfigure[\newline$\ln(C_\textrm{SK})=-2.67 + 3.40\ln(\ln(1/\epsilon))$\newline$\ln(C_\textrm{On})=-0.49 + 1.29\ln(\ln(1/\epsilon))+\ln3$\newline$\ln(C_\textrm{Off})=-0.72 + 2.27\ln(\ln(1/\epsilon))+\ln3$\newline]{
    \includegraphics[width=6 cm]{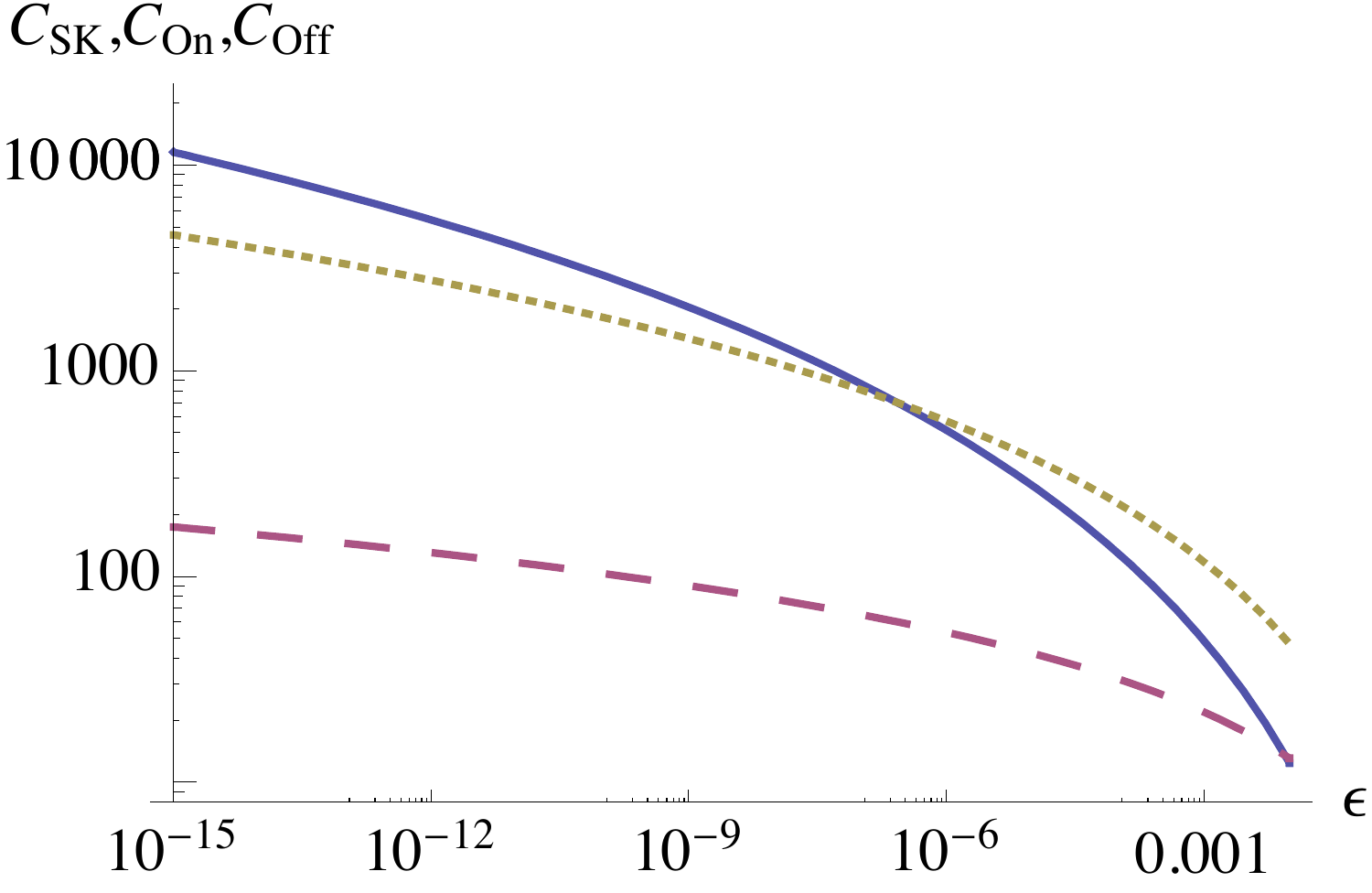}
    \label{fig:ComparisonSKU}
  }
  \caption{Full line: Cost of Solovay Kitaev decompostion (SKD) of (a) random $Z$-rotations and (b) random unitaries as a function of the precision $\epsilon$. Dotted line: Offline costs of (a) random $Z$-rotations and (b) random unitaries. We assume that, loosely speaking, a random unitary is the composition of three random rotations, hence the additional factor of three. Dashed line: Online cost of (a) random $Z$-rotations and (b) random unitaries. (a) For practical values of $\epsilon$, the online cost is significantly smaller than SK. The offline cost is lower than SK when $\epsilon\leq8.71\times10^{-4}$. (b) Again, for practical values of $\epsilon$, the online cost is significantly smaller than SK. The offline cost is lower than SK when $\epsilon\leq2.67\times10^{-7}$.}\label{fig:ComparisonsSK}
\end{figure}

\begin{figure}
  \subfigure[\newline$\ln(C'^Z_\textrm{SK})=-4.88 + 4.41\ln(\ln(1/\epsilon))$\newline$\ln(C'^Z_\textrm{On})=-0.78 + 1.12\ln(\ln(1/\epsilon))$\newline$\ln(C'^Z_\textrm{Off})=0.54 + 1.75\ln(\ln(1/\epsilon))$\newline]{
    \includegraphics[width=6 cm]{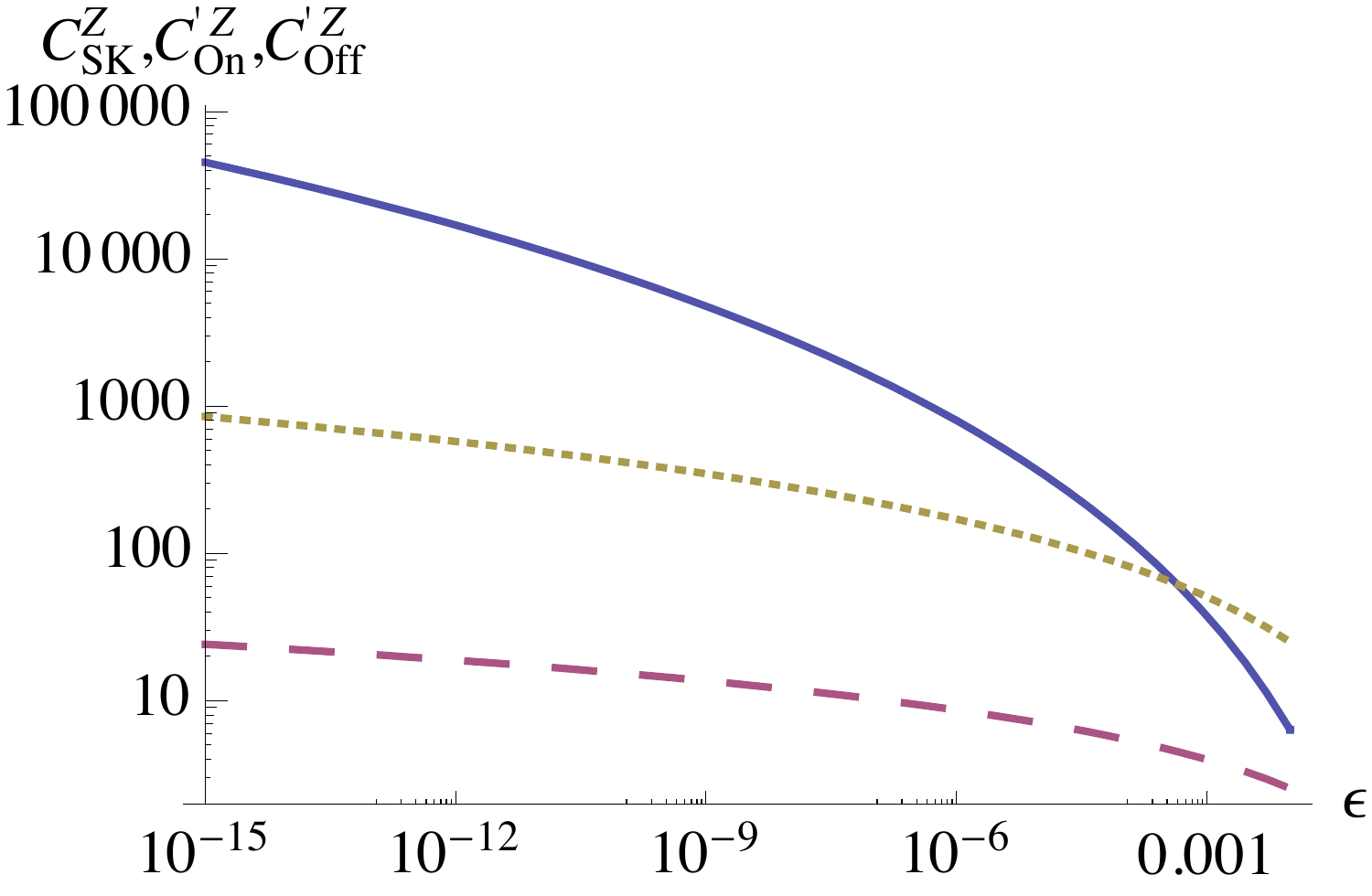}
    \label{fig:ComparisonSKMultiZ}
  }
  \subfigure[\newline$\ln(C'_\textrm{SK})=-2.67 + 3.40\ln(\ln(1/\epsilon))$\newline$\ln(C'_\textrm{On})=-0.78 + 1.12\ln(\ln(1/\epsilon))+\ln3$\newline$\ln(C'_\textrm{Off})=0.54 + 1.75\ln(\ln(1/\epsilon))+\ln3$\newline]{
    \includegraphics[width=6 cm]{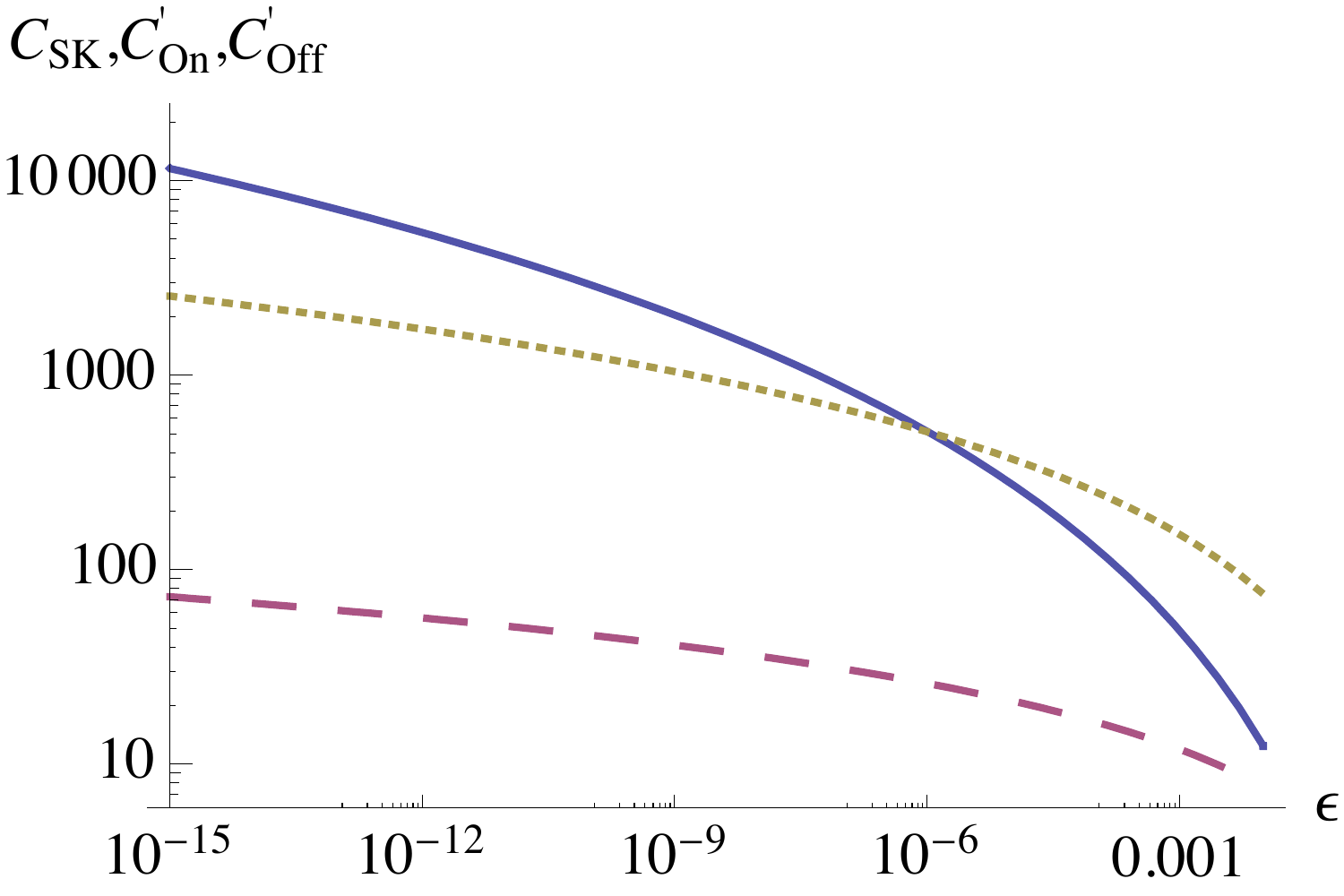}
    \label{fig:ComparisonSKMultiU}
  }
  \caption{Full line: Cost of Solovay Kitaev decompostion (SKD) of (a) random $Z$-rotations and (b) random unitaries as a function of the precision $\epsilon$. Dotted line: Offline costs, using $\{\ket H,\ket{\psi^0},\ket{\psi^1},\ket{\psi^2}\}$ as initial resources, of (a) random $Z$-rotations and (b) random unitaries. Dashed line: Online cost, using $\{\ket H,\ket{\psi^0},\ket{\psi^1},\ket{\psi^2}\}$ as initial resources, of (a) random $Z$-rotations and (b) random unitaries. (a) For practical values of $\epsilon$, the online cost is significantly smaller than SK. The offline cost is lower than SK when $\epsilon\leq4.41\times10^{-4}$. (b) Again, for practical values of $\epsilon$, the online cost is significantly smaller than SK. The offline cost is lower than SK when $\epsilon\leq1.03\times10^{-6}$.}\label{fig:ComparisonsSKMulti}
 \end{figure}

\begin{table}
  \centering
  \begin{tabular}{|c|c|l|l|l|}
    \hline
    $\theta$ & $C$ & $\epsilon=10^{-4}$ & $\epsilon=10^{-8}$ & $\epsilon=10^{-12}$\\
    \hline\hline
    $\pi/16$ & $C_{SK}$ & 43.83 & 2646 & 29120\\
    &$C_\textrm{on}$ & 10.20 & 24.52 & 41.95\\
    &$C'_\textrm{on}$ & 5.88 & 12.48 & 19.38\\
    &$C_\textrm{off}$ & 73.06 & 349.8 & 874.4\\
    &$C'_\textrm{off}$ & 98.29 & 306.1 & 595.0\\
    \hline\hline
    $\pi/128$ &$C_{SK}$ & 53.84 & 2879 & 29530\\
    &$C_\textrm{on}$ & 5.47 & 18.96 & 39.27\\
    &$C'_\textrm{on}$ & 3.32 & 9.27 & 16.91\\
    &$C_\textrm{off}$ & 49.18 & 313.0 & 923.9\\
    &$C'_\textrm{off}$ & 52.60 & 234.1 & 560.8\\
    \hline\hline
    $\pi/1024$ &$C_\textrm{SK}$ & 128.1 & 2594 & 15075\\
    &$C_\textrm{on}$ & 7.99 & 23.08 & 42.93\\
    &$C'_\textrm{on}$ & 3.00 & 8.37 & 15.23\\
    &$C_\textrm{off}$ & 77.42 & 381.3 & 969.1\\
    &$C'_\textrm{off}$ & 65.75 & 245.5 & 530.7\\
    \hline
  \end{tabular}
  \caption{$C_\textrm{on}$ and $C_\textrm{off}$ are respectively the online and offline costs to implement the $Z$-rotation by angle $\theta$ using only $\ket H$ states, to precision $\epsilon$. $C'_\textrm{on}$ and $C'_\textrm{off}$ refer to the costs for the version of the scheme thats uses $\{\ket H, \ket{\psi^0_0},\ket{\psi^1_0},\ket{\psi^2_0}\}$ as initial resource states. $C_\textrm{SK}$ refers to the extrapolated cost to implement these gates using the results from \cite{BS12}.}\label{tab:examplesRot}
\end{table}

Table \ref{tab:examplesRot} lists the expected costs. $C_\textrm{on}$ and $C_\textrm{off}$ are respectively the online and offline costs to implement the gate using only $\ket H$ states. $C'_\textrm{on}$ and $C'_\textrm{off}$ refer to the costs for the version of the scheme thats uses $\{\ket H, \ket{\psi^0_0},\ket{\psi^1_0},\ket{\psi^2_0}\}$ as initial resource states. $C_\textrm{SK}$ refers to the extrapolated cost to implement these gates using the results from \cite{BS12}. This extrapolated cost averages over all unitaries ($c\sim3.40$). This is optimistic since the results of \cite{BS12} suggest that $Z$-rotations are actually harder to implement ($c\sim4.3$).  Note that in theory, the cost of this algorithm is $O(\log ^c(1/\epsilon)$, where $c=3.97$ \cite{DN05}.

In all cases, the online cost is minimal when our proposed scheme enhanced by $\{\ket{\psi^0_0},\ket{\psi^1_0},\ket{\psi^2_0}\}$ is used.
For rougher precision, e.g., $10^{-4}$, the offline cost might be such that the total cost is still minimal for the Solovay-Kitaev implementation of \cite{BS12}.
For finer precision, e.g., $10^{-8}$ or $10^{-12}$, our proposed protocol becomes very advantageous, since the cost of the Solovay-Kitaev decomposition becomes prohibitive.

We also compare the average behavior of the different schemes as Figs.~\ref{fig:ComparisonsSK} and \ref{fig:ComparisonsSKMulti} illustrate.
We first start by noting that, loosely speaking, a random unitary is composed of three random rotations, such that the curves presented previously in Fig.~\ref{fig:ComparisionHMulti} must be shifted by $\ln 3$ in general.
Fig.~\ref{fig:ComparisonsSK} plots the fit for the Solovay-Kitaev decomposition (solid line), the online cost (dashed) and offline cost (dotted).
For all practical accuracies, the online cost of our proposed scheme is consistently the smallest.
However, the offline cost becomes advantageous when $\epsilon<8.71\times10^{-4}$ for $Z$-rotations and $\epsilon<2.67\times10^{-7}$ for random unitaries.
Fig.~\ref{fig:ComparisonsSKMulti} plots the same for the scheme with additional initial resource states.
Similarly, the offline cost becomes advantageous when $\epsilon<4.41\times10^{-4}$ for $Z$-rotations and $\epsilon<1.03\times10^{-6}$ for random unitaries.

\section{Conclusion}

We have proposed an alternative protocol to Solovay-Kitaev decomposition that results in significantly smaller resource costs, in both the number of required resource states and the depth of the circuit.
We have shown a significant improvement on average in the value of $c$, and in many cases the number of distilled states and rotations required to implement a single-qubit unitary gate are reduced.
Another advantage of our protocol is that the number of resources required is a ``smoother" function of accuracy,
whereas Solovay-Kitaev decomposition is step-like in nature because of the recursion process used in practice.
However, note that our protocols and Solovay-Kitaev decomposition are not exclusive.
It might be that some unitaries are better implemented using Solovay-Kitaev decomposition, while our scheme is better suited for $Z$-rotations, which occur, among other algorithms, in the quantum Fourier transform.

As future research, there are likely a variety of other circuits that enable other ``ladders" of states.
One natural extension would be to use the $SH$ eigenstates distilled using the protocols of \cite{BK05,BH12}.
Another extension would be to perform a systematic study of ``small" Clifford circuits.
Finally, we note that implementing a rotation by choosing the state which results in an angle closest to the target angle is a simple way of achieving our goal, but it is surely suboptimal.
An important research direction would be to optimize the sequence of angles required to implement the desired rotation.

\section{Acknowledgements}

We thank Alex Bocharov for many useful discussions.

\end{document}